\documentclass[amsmat,amssymb,amsfonts,aps,prl,twocolumn,showpacs]{revtex4}
\usepackage{graphicx}
\usepackage{dcolumn}
\usepackage{bm}
\newcommand{\beq}{\begin{equation}}  
\newcommand{\eeq}{\end{equation}}  
\newcommand{\beqa}{\begin{eqnarray}}  
\newcommand{\eeqa}{\end{eqnarray}}

\begin{document}

\title{Storage of classical information in quantum spins}

\author{F. Delgado$^{(1,2)}$  and J. Fern\'andez-Rossier$^{(1,2)}$}
\affiliation{$^{(1)}$ International Iberian Nanotechnology Laboratory (INL),
Av. Mestre Jos\'e Veiga, 4715-330 Braga, Portugal
\\ $^{(2)}$ Departamento de F\'isica Aplicada, Universidad de Alicante, 03690 San Vicente del Raspeig, Spain}

\begin{abstract}
Digital magnetic recording is based on the storage  of a bit of information in the orientation of a magnetic system with two stable ground states.   Here we address two fundamental  problems that arise when this is done on  a quantized spin:   
quantum spin tunneling  and back-action of the readout process. 
  We show that  fundamental differences exist between integer and semi-integer spins when it comes to both, read and record  classical information in a quantized spin.   Our findings imply  fundamental limits to the miniaturization of magnetic bits and  are relevant to recent experiments where
spin polarized scanning tunneling microscope   reads and records  a classical bit  in the spin orientation of a single magnetic  atom.
 \end{abstract}

 \maketitle


Recent experimental breakthroughs have laid the foundations for atomic-scale data storage, showing the capability to read and manipulate the spin of a single magnetic atom with a spin polarized scanning tunneling microscope (SP-STM)\cite{Wiesendanger_revmod_2009,Loth_Bergmann_natphys_2010}.  Read out is based on tunneling magnetoresistance at the atomic scale\cite{Wiesendanger_revmod_2009}: for a fixed current spin polarization in the tip,  the resistance is higher when the magnetic adatom spin is antiparallel to it.
 Spin manipulation is based on spin-transfer torque at the atomic scale\cite{Loth_Bergmann_natphys_2010,Delgado_Palacios_prl_2010}: angular momentum is transferred from the spin-polarized tunneling electrons to the magnetic atom. We are specially interested in cases where 
the magnetic atom is weakly coupled to the conducting substrate, which can be done thanks to a  monoatomic layer of insulating material\cite{Hirjibehedin_Lin_Science_2007}.  
As a result, the spin of the magnetic atom is quantized, and can be described 
by a single spin Hamiltonian,  as revealed by inelastic electron tunneling spectroscopy (IETS)\cite{Hirjibehedin_Lin_Science_2007,Loth_Bergmann_natphys_2010,Otte_Ternes_natphys_2008}, identical to that of single molecule magnets\cite{Gatteschi_Sessoli_book_2006}.

In this letter we address two fundamental questions that arise when considering  magnetic recording in the quantum limit, i.e.,  the storage and readout of a classical bit of information
 on a quantum spin. 
 First, 
what is the role played by spin-parity in the readout and control operations of a quantized spin?. 
The two physically different ground states required  to encode the two logical states of a bit
appear only in the case of semi-integer spin\cite{Loss_DiVincenzo_prl_1992} for which quantum spin tunneling\cite{Wernsdorfer_Sessoli_science_1999,Gatteschi_Sessoli_book_2006} is forbidden.  We also show that  current induced single atom spin switching is only possible for semi-integer spin.  The second question is how can the magnetoresistive single spin readout be performed without disturbing the spin state? Here   we study  the problem of the back-action, akin to the quantum non-demolition\cite{Braginsky_Khalili_revmodephys_1996} problem on a decohered qubit.

The physical system of interest consist on a magnetic atom with  quantized spin $S$\cite{Hirjibehedin_Lin_Science_2007,Otte_Ternes_natphys_2008,Loth_Bergmann_natphys_2010}.
 The magnetic atom is probed and controlled by a SP-STM. The quantized spin of an atomic scale nano-magnet  on a surface can be described with a single ion Hamiltonian\cite{Hirjibehedin_Lin_Science_2007,Loth_Bergmann_natphys_2010}: 
\begin{equation}
{\cal H}_{\rm Spin}= D \hat{S}_z^{2} + E(\hat{S}_x^{2}-\hat{S}_y^{2})+g\mu_B \vec{\hat{S}}.\vec{ B},
\label{Hspin}
\end{equation}
where $D$ and $E$ define the uniaxial and in-plane magnetic anisotropy. The eigenvalues and eigenfunctions of (\ref{Hspin}) are denoted by $E_M$ and $|M\rangle$, respectively.  The above Hamiltonian accounts for the measured 
IETS in $S=1$ Fe Phthalocyanine\cite{Tsukahara_Noto_prl_2009}, $S=3/2$  Cobalt 
adatoms\cite{Otte_Ternes_natphys_2008}, $S=2$  Fe adatoms and $S=5/2$ Mn adatoms\cite{Hirjibehedin_Lin_Science_2007}.
 
 The Hamiltonian of the total system features a single ion Hamiltonian exchange-coupled to the transport electrons\cite{Delgado_Palacios_prl_2010,Novaes_Lorente_prb_2010,Fransson_nanolett_2009},
${\cal H}= {\cal H}_{\rm T} + {\cal H}_{\rm S} + {\cal H}_{\rm Spin}+ {\cal V}$, where 
${\cal H}_{\rm T}+{\cal H}_{\rm S}=\sum_{\lambda,\sigma} \epsilon_{\sigma}(\lambda)  
c^{\dagger}_{\lambda,\sigma}c_{\lambda,\sigma}\;$ describes  the tip and surface electrodes, 
with   quantum numbers  $\lambda\equiv (\vec{k},\eta)$ and $\sigma$,  the momentum, electrode ($\eta=T,S$) and spin projection $\sigma$ along the tip polarization axis. We assume a spin polarized tip with polarization $\vec{{\cal P}}_T$ and a spin-unpolarized substrate $S$.
The  ${\cal V}$ term introduces interactions between tip, surface and the magnetic atom: 
\begin{equation}
{\cal V}=
\sum_{\lambda,\lambda',\sigma,\sigma'} \left(
{\cal T}^{(0)}_{\lambda,\lambda'}\frac{\delta_{\sigma\sigma'}}{2}+
{\cal T}_{\lambda,\lambda'}\vec{S}.\frac{\vec{\sigma}_{\sigma\sigma'}}{2}
\right)c^{\dagger}_{\lambda,\sigma} c_{\lambda'\sigma'}, 
\label{HTUN}
\end{equation}
  with $\vec{\sigma}$ the Pauli matrices vector and $\vec{S}$ the magnetic atom spin.
Equation (\ref{HTUN})  describes both 
 spin-independent tunneling, described by the ${\cal T}_{\lambda,\lambda'}^{(0)}$ term,
as well as  spin dependent processes, described by the ${\cal T}_{\lambda,\lambda'}$ term,  where  carriers can either remain in the same electrode, providing the most efficient atomic-spin relaxation channel, or switch sides, which gives rise to spin-dependent tunneling current.
Neglecting the momentum dependence and considering the initial and final electrode of the scattering events, there are 6 non-equivalent exchange integrals.
We describe them as ${\cal T}_{\lambda\lambda'}^{(0)}=v_\eta v_{\eta'}T_0$  and ${\cal T}_{\lambda,\lambda'}= T_J v_{\eta}v_{\eta'}$, with 
 the spinless $T_0$ and spinfull $T_J$ tunneling matrix elements, and two dimensionless scaling parameters that describe the strength of the tip-atom and surface-atom single-particle hoppings, $v_T$ and $v_S$.

  The effect of ${\cal V}$ on the energy levels is considered weak in the sense that it can be described within lowest order Fermi golden rule. 
 We  assume that the correlation time of the reservoirs formed by the electron gases at the tip and surface is short enough so that non-markovian effects are negligible\cite{Cohen_Grynberg_book_1998}. 
The dissipative dynamics of the atomic spin described by ${\cal H}_{Spin}$, under the influence of the dissipative coupling to the tip and substrate, is described  in terms of a Bloch-Redfield (BR)  master equation
 in which the coupling to the reservoirs is included up to second order in the coupling ${\cal V}$:
$\partial_t \hat{\rho}=-\frac{i}{\hbar}[{\cal H}_{Spin},\hat{\rho}]+{\cal L}\hat{\rho}$, with ${\cal L}$ the Liouvillian that accounts for the Kondo coupling  ${\cal V}$\cite{Cohen_Grynberg_book_1998}.  This equation describes the evolution of the diagonal terms in the density matrix, the occupations $P_M\equiv \rho_{M,M}$,  as well as the off-diagonal terms or coherences, $\rho_{M,M'}$.   In steady state, the density matrix $\rho_{M,M'}$ described by the BR master equation does not contain coherences, and only the diagonal terms $P_M$ survive\cite{Cohen_Grynberg_book_1998}.

The relevant scattering rates can written in terms of
\begin{equation}
 \gamma_{\eta,\eta'}^{aa'}(\epsilon) =  T_a T_{a'} \rho_{\eta}\rho_{\eta'} v_{\eta}^2v_{\eta'}^2\frac{\pi \epsilon}{2\hbar},
\end{equation}
where $\epsilon$ is some energy scale relevant for the process in question, $a$ can be $0$ or $J$, and $\rho_\eta$ is the density of states at the Fermi energy in electrode $\eta$.
 The  elastic conductance has a contribution coming from the spinless tunneling, $g_0 \equiv 2e^2  \frac{\partial \gamma_{TS}^{00}(\epsilon)}{\partial\epsilon}$, which plays no role in the remainder of the manuscript ($e$ is the (negative) electron charge).  From the experimental linear conductance we get that $\gamma_{TS}^{00}({\rm 1meV})=I/e\sim 0.1-5.$GHz\cite{Loth_Bergmann_natphys_2010}.
 
The spin-readout is based on a second  contribution to the elastic conductance coming from  elastic exchange between transport electrons and the spin of the magnetic atom that gives rise to spin-valve term in the total conductance\cite{Delgado_Palacios_prl_2010}:
\begin{eqnarray}
&&G_{\rm el}(V) \approx g_0 \left[1+  2\frac{T_J}{T_0}\langle \vec{S}\rangle. \vec{{\cal P}}_T\right],
\label{ielast-mag}
\end{eqnarray}
where 
$\langle \vec{S}\rangle$ is the expectation value of the electronic spin:
\beqa
\langle \vec{S} \rangle=\sum_M P_M(V) \langle M|\vec{S}|M\rangle.
\label{averagespin}
\eeqa
Thus, for finite tip polarization, the conductance is sensitive to the expected value of the atomic magnetic moment along $z$.  Thus, if the quantum spin can be in  two different spin states at zero applied field,
ideally with  $\langle M| \vec{S}|M \rangle$  parallel and antiparallel to the tip moment, 
  then a magnetoresistive  readout of a classical bit of information on a quantum spin is possible. 
  
  We now discuss the necessary conditions for the existence two  ground states. 
First,  $D$ should be negative.  To see this,  we  consider first the  idealized case of a quantum spin with $E=0$. The energy levels are $E_M= D S_z^2$, with $S_z=\pm S$, $\pm (S-1)$ ...    If $D$ is positive, the ground states doublet would have $S_z=\pm 1/2$ for semi-integer spin,  which can give rise to Kondo effect\cite{Otte_Ternes_natphys_2008},  or $S_z=0$ for integer spin. In both cases the magnetic moment is zero. 
Second, the spin should be integer.   
Kramer's theorem\cite{Kramers_paa_1930} states that, at zero field and with $E\neq0$,  integer spin systems have non-degenerate spectrum, but semi-integer spins have, at least, a twofold degeneracy.   These zero-field splittings can be interpreted in terms of quantum spin tunneling, which is suppressed for semi-integer $S$\cite{Loss_DiVincenzo_prl_1992}. 
Thus, the $E$ term  splits all the doublets of the $E=0$ spectrum only for integer $S$. 
 Zero field splitting for integer spins  has  a very important consequence, which derives from the following general result.  For zero applied magnetic field, the matrix elements $\langle M|\vec{S}|M\rangle$ are zero for every non-degenerate eigenstate of ${\cal H}_{Spin}$\cite{Klein_ajp_1951}.  Thus,  from Eq. (\ref{averagespin}) we get that, at zero applied field, it is impossible to have a net magnetic moment for integer spins. 
In contrast, for semi-integer $S$, an arbitrary small magnetic field along an arbitrary direction $\hat{\Omega}$ will choose  between the two ground states $g_+$ and $g_-$,  resulting in $\langle g_{\pm}|\vec{S}\cdot \hat{\Omega}|g_\pm\rangle\neq0$. 
These two states provide the physical realization of the two  logical states of the classical bit.

\begin{figure}
\includegraphics[width=0.99\linewidth,angle=0]{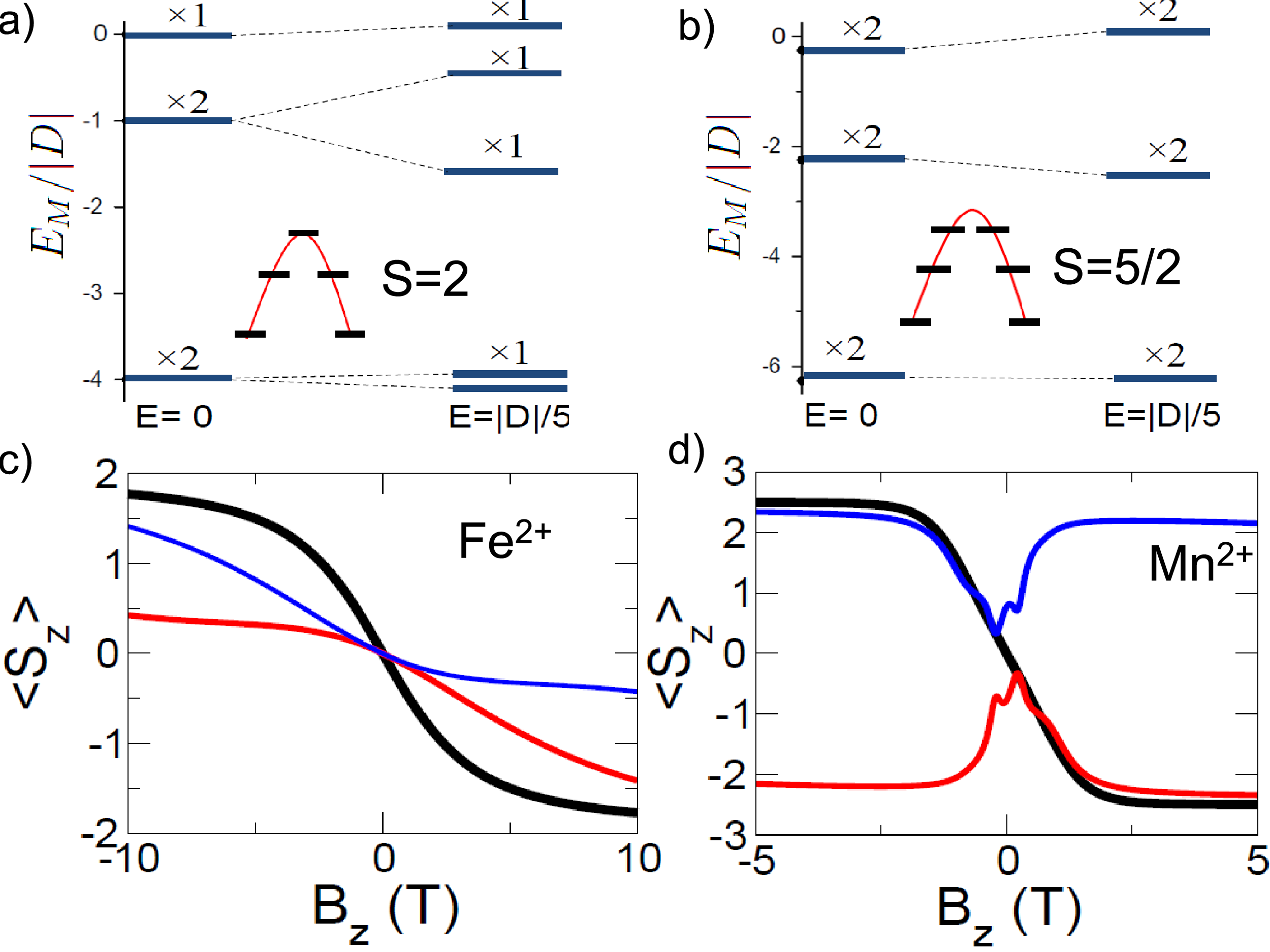} 
\caption{(Color online) Schematic evolution of the energy spectrum and degeneracies versus $E$ for: a) integer spin $S=2$ and b) half-integer spin $S=5/2$. Magnetization curves of the Fe c) and Mn d) adatoms probed with a spin polarized tip with ${\bf {\cal P}}_T=0.74$ for three applied bias: $V=0$ (thick black line), $V=-10$ meV (red line), and $V=+10$ meV (blue line). Here $T=0.5$K, $T_J/T_0=0.5$, $v_\eta=1$, and $\rho_T=\rho_S$.
}
\label{fig1}
\end{figure}

IETS confirms  this scenario for  Fe $(S=2)$  and Mn $(S=5/2)$ on Cu$_2$N \cite{Hirjibehedin_Lin_Science_2007}. 
In both cases, $D$ is negative ($D_{Fe}=-1.55$ meV, $D_{Mn}=-39\mu$eV). However, 
in the case of  Fe, there is a single ground state due to quantum spin tunneling  induced by $E$, with  a null  average magnetization.  In contrast, for Mn, with $S=5/2$  
 the in-plane anisotropy does not lift the degeneracy of the ground state doublet for the Mn, see Fig.~\ref{fig1}a),b) 
  
\begin{figure}
[t]
\includegraphics[width=0.99\linewidth,angle=0]{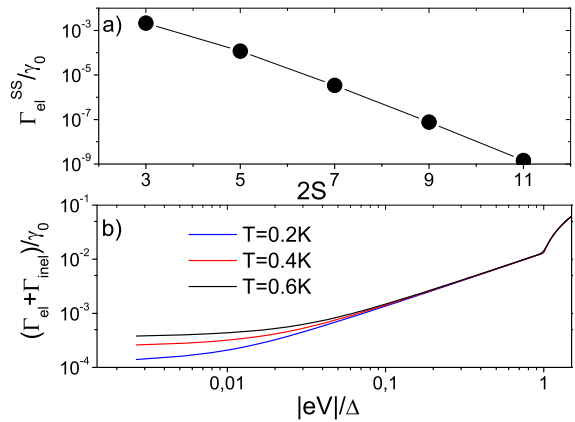} 
\vspace{-0.9cm}
\caption{(Color online) a) Substrate mediated elastic spin relaxation rate $\Gamma_{ela}^{SS}$ in units of $\gamma_{0}\equiv \gamma_{S,S}^{JJ}(1{\rm meV})$
for an ideal half-integer spin systems with $D=-5|E|=-1$meV and $T=0.4$K. b) Total relaxation rate $(\Gamma_{inel}+\Gamma_{ela})$ for $S=5/2$ at three different temperatures. Here ${\cal P}_T=-1$, $v_S=1$, $v_T=0.7$, $T_J/T_0=0.5$ and $\rho_S^2 T_J^2=0.01$.
}
\label{fig2}
\end{figure}

The storage of information in $D<0$ semi-integer spins is limited by spin relaxation, originated both  by elastic and inelastic processes. The later, addressed below,  are exponentially suppressed when both bias and temperature are smaller than the excitation energy.
In contrast, the rate of elastic scattering between the two ground states  $g_{\pm}$ due to coupling to the   substrate,
 for semi-integer $S$  is given by
  \beqa
\Gamma_{\rm el}=
\gamma_{S,S}^{JJ}(k_BT)
\sum_{a=x,y,z} |\langle g_-|S_a|g_+\rangle|^2,
\label{direct}
\eeqa
with $k_B$ the Boltzmann constant. The wave functions $|g_{\pm}\rangle $ satisfy
\begin{equation}
|g_{\pm}\rangle  
\propto\left(|\pm S\rangle + \sum_n c_n \left(\frac{E}{D}\right)^{2n} |\pm S \mp 2n\rangle\right),
\end{equation}
where  $n=1,2,...,S-\frac{1}{2}$, and $c_n$ are dimensionless numbers of order 1. We see that in-plane anisotropy enables the exchange assisted elastic spin flip\cite{Romeike_Wegewijs_prl_2006II}  $|\langle g_-|S_a|g_+\rangle|^2\propto (E/D)^{2S-1}$.  Thus, elastic population scattering is suppressed as either $S$ or $D/E$ increase. Numerical calculation (see Fig. \ref{fig2}a) yield lifetimes in the range of microseconds for $S=5/2$ and $|D|=5E$ at 0.4K. 

Importantly,  the coupling of the atomic spin to the conduction electrons kills the coherence between the two ground states $g_\pm$, 
which   satisfies the equation $\partial_t \rho_{g_+,g_-}=-\Gamma_{g_+,g_-} \rho_{g_+,g_-}$. The decoherence rate   $\Gamma_{g_+,g_-}$ contains contributions from both the  non adiabatic terms,  that imply population scattering, like in Eq. (\ref{direct}),  and adiabatic terms,  which do not involve energy exchange with the reservoir\cite{Cohen_Grynberg_book_1998}.  The substrate mediated rate for the adiabatic term reads as:
\begin{equation}
\Gamma^{ad}_{g_+,g_-} =
\frac{\gamma_{SS}^{JJ}(k_B T)}{4}
 \Big|\langle g_+| S_z|g_+\rangle-\langle g_-| S_z|g_-\rangle\Big|^2.
  \label{decoh}
\end{equation}
 Thus, coupling to the electronic environment kills quantum coherence more efficiently in states with opposite magnetic moments,
 acting as a which path detector\cite{Stern_Aharonov_pra_1990} and  favoring the magnetoresistive readout.   
  To leading order in $E/|D|$,  we have   $\Gamma_{g_+,g_-}^{ad}\simeq \gamma_{SS}^{JJ}(k_B T) S^2$.  In contrast with population scattering, decoherence rate increases for larger $S$ and is independent of $E/|D|$.   Thus, larger $S$ favors spin memory but kills quantum effects. 
The ratio of the adiabatic decoherence rate, Eq. (\ref{decoh}), and the elastic population scattering, Eq. (\ref{direct}), reads 
$S^2 \left(|D|/E\right)^{2S-1}$, which is above $10^3$ for Mn in Cu$_2$N.  

 We now turn our attention to  the effect of parity on the process of magnetic recording, 
based on atomic scale spin transfer torque, which has only been studied for semi-integer spins so far\cite{Loth_Bergmann_natphys_2010,Delgado_Palacios_prl_2010}. Current  flowing through the spin-polarized tip, 
 transfers angular momentum to the atomic spin.  When the transfer rate exceeds the spin relaxation rate, the spin is driven out of equilibrium. In the case of semi-integer $S$ at zero applied magnetic field,  
this can result in the occupation of one of the two decohered ground states, $g_\pm$,  and the depletion of the other, giving rise to a net magnetic moment $\langle S_z\rangle$, according to Eq. (\ref{averagespin}).   The population transfer takes place mainly through inelastic excitation of the spin from the ground state doublet $S_z=\pm S$  to the first excited doublet, via spin-flip exchange.   The transition rate where an $\uparrow$ (majority) electron from the tip 
spin-flips and goes to the surface reads (positive applied voltage in our sign convention)\cite{Delgado_Rossier_prb_2010}:
\begin{equation}
\Gamma_{\rm inel} \approx 
\gamma^{JJ}_{TS}(|\Delta+eV|)
|\langle g_+|S^+|x_+\rangle|^2, 
\label{inel}
\end{equation}
where we have assumed that $|eV| \gg \Delta,\;k_BT$ while $|x_+\rangle$ refers to the excited state connected to $g_+$. In fact, the efficiency of the process is greatly enhanced when either bias or temperature are higher than the inelastic excitation energy, $\Delta \simeq (2S-1)|D|$ for half-integer spin $S$.

 In the case of integer spins, inelastic excitations also transfer population between the two tunnel-split ground states but, as the expectation value of the magnetic moment in Eq. (\ref{averagespin}) at zero applied field in both states is null, $\langle S_z\rangle=0$, no matter which non-equilibrium distribution is achieved.
In Fig. \ref{fig1}c),d) we plot   $\langle S_z\rangle$, defined in Eq. (\ref{averagespin}), as a function of a magnetic field for 3 situations: zero bias, +$10$meV and -$10$meV, for both Fe and Mn on Cu$_2$N with finite tip polarization. 
At zero bias, we obtain the equilibrium  Brillouin curve\cite{Ashcroft_Mermin_book_1976}. 
At finite bias  spin transfer favors spin alignment parallel ($V<0$) or antiparallel ($V>0$) to the magnetic moment of the tip. 
The striking difference between integer and semi-integer  spin is apparent in the figure. For integer spin, the magnetic moment is always null at zero field and the effect of bias is to heat the atomic spin decreasing
 the absolute value of $\langle S_z \rangle$ with respect to the zero bias case. For semi-integer spin,  the atomic spin takes a bias dependent value at zero field. 
Thus, we find that current driven control of the magnetic moment of a single spin is only possible for semi-integer $S$.

We now address the problem of back-action and the conditions under which a SP-STM can perform the quantized spin readout without perturbing the atomic spin state, avoiding the loss of the classical information.
 In other words, we look for 
a quantum non-demolition measurement\cite{Braginsky_Khalili_revmodephys_1996} of the atomic spin using  SP-STM, with the caveat that the atomic spin is decohered.  The magnetoresistive read-out [Eq. (\ref{ielast-mag})]  is made possible by  the tunneling exchange coupling between the quantum spin and the transport electrons. 
Specifically, 
it is based on the non-spin flip or Ising coupling, $S_z\sigma_z$, which does not flip the atomic spin.  However,  if tunneling exchange is spin-rotational invariant, Eq. (\ref{HTUN}), the Ising term goes together with 
   the flip-flop terms, $S^+\sigma^-+S^-\sigma^+$, which induces atomic spin scattering with the selection rule $\Delta S_z=\pm 1$ and are responsible of the recording (spin-transfer torque).   Thus, as in many other instances, 
the reading mechanism entails some degree of back-action on the probed system.  The back-action occurs via inelastic spin-flip events, whose  rate $\Gamma_{\rm inel}$  takes off  when either bias or temperature provides the excitation energy, and  the elastic spin-tunneling assisted spin-flip, whose  rate $\Gamma_{el}$ depends only on $k_BT$, see Fig.~\ref{fig2}.

The condition for non-demolition readout is that the measuring time $\tau$ is significantly shorter than the spin-lifetime,
     $\tau^{-1}\gg (\Gamma_{\rm inel}+  \Gamma_{\rm el})$.  Regardless of the instrumentation,  
     the measuring time has a fundamental limit  given by the condition that shot noise $\delta I$ should be smaller than the 
     current contrast,  $\Delta I= \Delta G V$.   For Poissonian noise,  we have $\delta I= \sqrt{\frac{e}{\tau}\overline{I}}$, where $\overline{I}$ is the average current measured during $\tau$.  If we define the average time for a single electron passage , $\tau_e=e/{\bar I}$, 
 then, the limit imposed by shot noise is $\tau\gg\tau_e$. In other words, many tunneling events are necessary to perform the magneto-resistive single spin readout.
     
     Current experiments are done with  $\overline{I}$ in the range of nA, which yields     $\tau_e\sim 0.2$ ns, so that  the  measuring time is bound by below, due to  shot noise, by 1ns. State of the art instrumentation requires much larger measuring times.  For instance,  the use of lock-in introduces a more stringent bound to $\tau$, in the range of 1$\mu$s-1ms\cite{Sloan_jphyscm_2010,Guo_Hihath_nanolett_2011}.  In Fig. \ref{fig2}a) we plot the
substrate mediated elastic spin relaxation rate $\Gamma_{ela}^{SS}$ for and ideal spin system with $D=-5|E|=-1$meV and $T=0.4$K. This relaxation time grows exponentially with the spin $S$. For an experimentally sensible zero bias conductance $G(0)\approx 0.01G_0$\cite{Loth_Bergmann_natphys_2010} ($\rho_S T_J=0.1$), relaxation time above 1$\mu$s can be found in system with $S\ge 5/2$ with excitation energies $\Delta \gtrsim  4$ meV at $T=0.4K$. The bias dependence of the total relaxation rate is shown in Fig.~\ref{fig2}b)\cite{Delgado_Rossier_prb_2010}. In order to realize a non-demolition measurement, the bias should be kept $|eV|\ll \Delta$. As shown in Fig.~\ref{fig2}b), relaxation rate in this regime is dominated by the substrate mediated processes, Eqs. (\ref{direct}).

In summary,  we have studied the limitiations imposed by quantum mechanics to the use of quantum spins to store classical bits of information. We have found that 
   classical information can be stored in quantum semi-integer spins, for which quantum spin tunneling is suppressed and quantum spin torque is possible, with uniaxial anisotropy $D<0$. The storage time is limited, when un-observed, 
by the elastic spin-flip rate (Eq. \ref{direct}). Magnetoresistive readout  induces additional spin scattering given by the rate (\ref{inel}).  Shot noise imposes a lower limit to the measuring time.  Increasing $S$, 
using for instance few atom ferromagnetically coupled  spin clusters,   rises dramatically both the elastic and back-action lifetimes,  as well as the decoherence rate facilitating  the  magnetic recording on a quantum spin.

This work was supported by MEC-Spain (MAT07-67845,  FIS2010-21883-C02-01, 
Grants  JCI-2008-01885 and  CONSOLIDER CSD2007-00010) and Generalitat Valenciana (ACOMP/2010/070). 
We acknowledge useful conversations with R. Aguado, G. Saenz and C. Untiedt.




\end{document}